# Super-Exchange Theory for Polyvalent Anion Magnets

Fang Zhang[1,2], You-Chao Kong[1], Rui Pang[3], Liang Hu[2], Peng-Lai Gong*[,2], Xing-Qiang Shi*[,2], Zi-Kang Tang*[,1]

[1] Institute of Applied Physics and Materials Engineering, University of Macau, Macau S.A.R., People's Republic of China.

[2] Department of Physics, Southern University of Science and Technology, Shenzhen 518000, People's Republic of China.

[3] International Laboratory for Quantum Functional Materials of Henan and School of Physics and Engineering, Zhengzhou University, Zhengzhou 450001, People's Republic of China.

**Corresponding Authors**

*E-mails:   gongpl@sustc.edu.cn (P. L. Gong)

shixq@sustc.edu.cn (X. Q. Shi)

zktang@umac.mo (Z. K. Tang)

# Abstract

The Goodenough-Kanamori-Anderson (GKA) rules have been widely applied for explaining the magnetic properties induced by super-exchange interaction. As conclusions of the super-exchange theory, they reveal the antiferromagnetic (ferromagnetic) ordering along with bond angle of 180º (90º) in the cation-anion-cation interaction path, in which the theory sets a pre-condition that the electronic states of cations in all paths are identical. We observed that the GKA rules are in fact not universal and even invalid to materials containing anions with different valence states, for example, the layered CrOCl crystal (with two valence states of anions: $O^{2-}$ and $Cl^-$). In this study, we propose an extended super-exchange theory (ESET) related to superposed electronic states of cation in a specific path. ESET is capable of predicting not only the sign and relative magnitude of magnetic exchange constants in different cation-anion-cation paths, but also the magnetic ground state. Through our proposed theory, we conclude that the magnetic ordering along with bond angle of 90º in Cr-Cl-Cr path is moderately antiferromagnetic and of 180º in Cr-O-Cr path is strongly ferromagnetic, which are opposite to the contents of GKA rules. Moreover, we clarify that monolayer CrOCl has antiferromagnetic ordering rather than ferromagnetic as reported recently. The reliability of ESET is verified via first-principles calculation and previous experimental report as well, and its universality is also demonstrated. Thus, our theory is powerful to predict the magnetic properties, which makes it possible to design new high Curie temperature two-dimensional semiconducting ferromagnets with polyvalent anion materials.

# 1. Introduction

The magnetism of matters, especially whose transition metal cations interacting indirectly through intervening nonmetal anions, which called super-exchange interaction (SEI), is one of the most intricate topics in condensed matter physics and chemistry. Back to 1934, *Kramers*[1] originally suggested the existence of SEI, when he dealt with the exchange interaction in paramagnetic salts based on perturbation theory. His theory was later employed by *Anderson*[2] to interpret the antiferromagnetic (AFM) property of MnO[3]. Sooner after, *Anderson* realized that the theory involves increasing uncertainties and complexities, as the exchange effect is in the third order of perturbation. Consequently, despite a triumph the theory had been achieved, *Anderson*[4] proposed a new theory of SEI from the view point of orbital symmetry relations. The general form of interaction strength was given, but its sign or magnitude can only be revealed under certain conditions. Nowadays, the deemed most satisfactory theory of SEI is developed over a period of time by *Goodenough*[5] and *Kanamori*[6]. Main conclusions are summarized in terms of *Goodenough-Kanamori-Anderson* (GKA) rules[7]: the SEI of a 180º bond angle in cation-anion-cation path is strongly AFM, whereas a 90º angle SEI favors ferromagnetic (FM) and much weaker.

The GKA rules have acquired success in explaining magnetic properties on an amount of two-dimensional (2D) and bulk materials, *e.g.* recently highlighted 2D materials of $CrX_3$ (X = F, Cl, Br, I)[8-10], $ABX_3$ (A = Mn, Cr; B = P, Si, Ge; X = S, Se, Te)[11], and bulk materials of $AX_2$ (A = V, Cr, Mn; X = Cl, Br, I)[12], $AO_2$ and $BCl_3$ (A = Mn, Cu, Ni; B = Co, Ni, Cu)[13-14], *etc.*. While isolated counterexamples (*e.g.* $MnCl_2$)[15] do exist, there is no doubt about the accuracy[16] of GKA rules, which is the reason for its wide application. However, as new magnetic materials are being discovered, an open question is arising: can GKA rules —— the semiempirical conclusions from super-exchange theory, always maintain their accuracy? The electronic state of cation is of great significance to analyze the magnetic ordering. In the traditional class of magnetic materials, every anion possesses equivalent valence state, like $CrCl_3$[17](-1 for $Cl^-$). As a consequence, the electronic state ($3d^3$ for $Cr^{3+}$) of cations in all super-exchange paths (i.e. different cation-anion-cation paths) are totally identical. However, for the new class of materials that comprise polyvalent anions, such as CrOCl (-2 for $O^{2-}$, -1 for $Cl^-$), this pre-condition is compromised. If we take cation as $3d^3$ like usual, the SEI in the path Cr-O-Cr (or Cr-Cl-Cr) related to both O and Cl atoms, which is not the real SEI in the specific path, but is combination of both. Therefore, super-exchange theory is no longer applicable for polyvalent anion materials. A new theory is badly needed.

In the present work, we take recently proposed 2D magnetic CrOCl[18] as an example to theoretically analyze the SEI of polyvalent anion materials (valence state -2 for $O^{2-}$, -1 for $Cl^-$), which we call extended super-exchange theory (ESET). ESET firstly gives the valid electronic state of cation in specific super-exchange path, namely superposed $\frac{1}{\sqrt{2}}(|d^3\rangle + |d^4\rangle)$ state for path Cr-O-Cr and $\frac{1}{\sqrt{2}}(|d^4\rangle + |d^4\rangle)$ for path Cr-Cl-Cr. Then, we discuss how these cations interacting by using orbital symmetry relations between *d* and *p* orbitals acquired from quantum mechanics[19-20]. The results of ESET show that a 180º bond angle in the interaction path Cr-O-Cr favors strongly FM configuration and 90º bond angle in the interaction path Cr-Cl-Cr is moderately AFM ordering. Through our analytical results, the sign and relative magnitude of magnetic exchange constants in Heisenberg Spin Hamiltonian model[21] are given. In addition, the overall magnetic ground states of monolayer and bulk CrOCl are predicted, both revealed to be AFM. Our theory is then confirmed by both first-principles calculation and previous bulk experiment[22]. Besides CrOCl, ESET is also applied to monolayer FeOCl with a different electronic state ($3d^5$ for $Fe^{3+}$[23], see Supporting Information) to validate its universality. Since strongly ferromagnetic SEI does exist according to our theory, a new route for designing high Curie temperature semiconducting ferromagnets is opened. Thus, ESET is significant as more polyvalent anion materials are being studied, *e.g.* magnetic Mxenes [24-25], many easily exfoliate 2D magnetic metals and semiconductors[26].

## 2. Calculation Details

Our calculations were carried out using the projector-augmented-wave (PAW) method[27] as implemented in Vienna *ab initio* simulation package (VASP)[28]. For bulk CrOCl, van der Waals (vdW) correction was considered by DFT-D3 method [29-30]. We used 2 × 2 × 1 supercell containing 20 atoms and plane-wave cutoff energy of 500 eV for all calculations (monolayer and bulk). A $\Gamma$- centered Monkhorst-Pack *k*-point mesh of 5 × 6 × 5 and 5 × 6 × 1 were employed for bulk and monolayer. A vacuum space of 20 Å along the *z*-direction was set for monolayer. The convergence criteria for the energy difference in electronic self-consistent loop were $10^{-5}$ eV and residual forces on ions were less than 0.01 eV/Å. To count the electron correlation effects of Cr *3d* orbital and obtain accurate electronic properties, we employed the screened hybrid HSE06[31-32] functional that usually performs much better than the DFT+U methods [33-35]. Further details about optimized lattice constants and total energies of layered CrOCl can be found in Supporting Information.

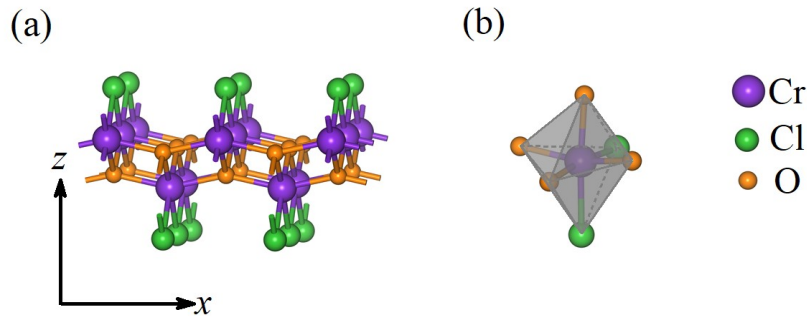

**Fig. 1. Crystal structures of CrOCl. (a)** Atomic structure from the front view. **(b)** Distorted octahedral structure of Cr atom. Cr atoms in monolayer have two in-plane layers, bulk crystal is AA stacking order.

# 3. Extended Super-exchange Theory

## 3.1 Electronic Configurations of Cation

The electronic states of cations, are of great importance in SEI[5]. In order to find that, we firstly define four states

$$|\psi_R\rangle = |\psi_o\rangle + |\psi_i\rangle + |\psi_c\rangle \tag{1}$$

where $|\psi_R\rangle$ is the real state that we take account in SEI. It has three components. The original electronic state of isolated atom $|\psi_o\rangle$ (e.g. $|\psi_o\rangle = |p^4\rangle$ for O atom), the ionic electronic state $|\psi_i\rangle$ changing from the original electronic state due to ionic character, and the covalent electronic state $|\psi_c\rangle$ changing from the original electronic state due to covalent character. Since Cl receives one electron, and O receives two, $|\psi_i\rangle_{Cl} = +|\varphi_e\rangle$, $|\psi_i\rangle_O = +2|\varphi_e\rangle$ where $|\varphi_e\rangle$ is the state of one electron. Because cations donate all electrons

$$\sum_{anions} |\psi_i\rangle + \sum_{cations} |\psi_i\rangle = 0 \tag{2}$$

which gives $|\psi_i\rangle_{Cr} = -3|\varphi_e\rangle$. The $|\psi_o\rangle + |\psi_i\rangle$ means the electronic state owning to total ionic bond (e.g. $|\psi_o\rangle + |\psi_i\rangle = |d^3\rangle$ for $Cr^{3+}$). However, ionic character is not alone, as covalent character always exists in real materials. In ESET, $|\psi_c\rangle$ for anions is

$$|\psi_c\rangle_{anion} = \begin{cases} 0 & (in\ super-exchange\ path) \\ -\dfrac{|\psi_i\rangle}{2} & (not\ in\ super-exchange\ path) \end{cases} \tag{3}$$

The difference between ESET and super-exchange theory is that ESET considers anions not in super-exchange path. For path Cr-Cl-Cr, the influence of O could be removed, based on the assumption that the polarized covalent bond between O and Cr are already formed, and vice versa. The population of electrons that anions lose due to covalent character, are assumed to be the half it receives due to fully ionic character. Eq. (2) is automatically suitable for $|\psi_c\rangle$, namely

$$\sum_{anions} |\psi_c\rangle + \sum_{cations} |\psi_c\rangle = 0 \tag{4}$$

In super-exchange theory, $|\psi_c\rangle_{anion}$ in super-exchange path equals to zero. For one-valent state materials, $|\psi_c\rangle_{cation}$ is then zero by Eq. (4), resulting in $|\psi_R\rangle_{cation} = |\psi_O\rangle + |\psi_i\rangle$ which can be directly deduced from chemical formula as one always did. However, for polyvalent anion materials, one specific path only related to one $|\psi_i\rangle$ state anion. There has non-zero $|\psi_i\rangle$ for anions not in super-exchange paths, which makes $|\psi_i\rangle$ for cations non-zero. So, Cr cannot take as $|d^3\rangle$ state in CrOCl material.

In path Cr-O-Cr, $|\psi_c\rangle_O = 0$, $|\psi_c\rangle_{Cl} = -\frac{1}{2}|\varphi_e\rangle$. According to Eq. (1) and (4), the electronic state of Cr atom is

$$|\psi_R\rangle_{Cr} = |d^3\rangle + \frac{1}{2}|\varphi_e\rangle = \frac{1}{\sqrt{2}}(|d^3\rangle + |d^4\rangle) \tag{5}$$

which means the state $|\psi_R\rangle_{Cr}$ in path Cr-O-Cr can regard as the superposition of states $|d^3\rangle$ and $|d^4\rangle$.

In path Cr-Cl-Cr, $|\psi_c\rangle_O = -|\varphi_e\rangle$, $|\psi_c\rangle_{Cl} = 0$, so

$$|\psi_R\rangle_{Cr} = |d^3\rangle + |\varphi_e\rangle = \frac{1}{\sqrt{2}}(|d^4\rangle + |d^4\rangle) \tag{6}$$

It should be noticed that the electronic state of cation in path Cr-O-Cr is a superposed state, described as the linear combination of states $|d^3\rangle$ and $|d^4\rangle$ for convenience of the following analysis. The interaction is always via super-exchange, not double-exchange[36].

**3.2 Orbital Symmetry Relations in Distorted Octahedral Crystal Field**

For super-exchange paths, the sign of exchange integral determines the magnetic ordering[6]. Here, we do not use the exchange integral method, but the basic principle of quantum mechanics [19-20], as the latter one is much simple and intuitive. In quantum mechanics, $d$ orbitals could be divided into five ($d_{xy}$, $d_{xz}$, $d_{yz}$, $d_{z^2}$, $d_{x^2-y^2}$) and $p$ orbitals into three ($p_x$, $p_y$, $p_z$) according to different magnetic quantum number. Among them, symmetry relations exist[13] between $d$ and $p$ orbitals, such as $d_{xz}$ is orthogonal to $p_z$, and $d_{z^2}$ is orthogonal to $p_x$. To avoid the cumbersome fifteen symmetry relations in total, we inherit the notation $t_{2g}$ and $e_g$ from crystal field theory[37] of octahedral structure, denoting $d$ orbitals in CrOCl as $t'_{2g}$ (contains $d_{xy}$, $d_{xz}$ and $d_{yz}$ orbitals) and $e'_g$ (contains $d_{z^2}$ and $d_{x^2-y^2}$ orbitals). If we choose $z$ as the main axis, the $p_z$ orbital that parallel to it is regarded as $p_\sigma$, the other two orbitals that perpendicular to it ($p_x$ and $p_y$) are called $p_\pi$. In this simplification, the fifteen orthogonality relations can be expressed as

$$\langle e'_g | p_\pi \rangle = \langle t'_{2g} | p_\sigma \rangle = 0$$
$$\langle e'_g | p_\sigma \rangle \gg \langle t'_{2g} | p_\pi \rangle \neq 0 \qquad (7)$$

The wavefunction has no overlap between orbitals $e'_g$ and $p_\pi$, $t'_{2g}$ and $p_\sigma$, indicating of forbidden electrons transition. Besides, the correlation of orbitals $e'_g$ and $p_\sigma$ is much stronger than $t'_{2g}$ and $p_\pi$.

Since the CrOCl crystal has a distorted octahedral structure (Figure 1b) like CrI$_3$[8] as the $d^3$ (Cr$^{3+}$) state in the latter one can write as[38] $(t_{2g})^3(e_g)^0$, it is reasonable to say in our CrOCl system, $d^3$ (Cr$^{3+}$) and $d^4$ (Cr$^{2+}$) states are $(t'_{2g})^3(e'_g)^0$, $(t'_{2g})^3(e'_g)^1$, respectively. More plausibility about this is discussed in Supporting Information.

### 3.3 Four Possible Super-Exchange Paths in CrOCl

Now, we determine the magnetic ordering along different paths. Apart from orbital symmetry relations discussed above, we also need Pauli exclusion principle (PEP) and Hund's rule (HR) to achieve magnetic ordering.

The superposed electronic state of Cr can be regarded as the linear combination of two states, we denote them as cation$_1$ and cation$_2$. Path cation$_1$-anion-cation$_2$ can align to two bond angles, 90º or 180º. If the angle is 180º, cation$_1$ and cation$_2$ share the same main axis, which means $p_\sigma$ orbital for anion in cation$_1$-anion bond is also $p_\sigma$ for cation$_2$-anion bond and so does $p_\pi$. For 90º angle, $p_\sigma$ orbital in cation$_1$-anion bond turns to $p_\pi$ in cation$_2$-anion bond and $p_\pi$ turns to $p_\sigma$, as the main axis of cation$_1$ and cation$_2$ are perpendicular. Combining with orbital symmetry relations, the SEI can only take place through $e'_g$-$p_\sigma/p_\pi$-$t'_{2g}$ for 90º, $e'_g$-$p_\sigma/p_\sigma$-$e'_g$ and $t'_{2g}$-$p_\pi/p_\pi$-$t'_{2g}$ for 180º. For simplicity, if the interaction of cation-anion is through orbital $e'_g$-$p_\sigma$, we call it $p_\sigma$ bond. For $t'_{2g}$-$p_\pi$ connection, denoted as $p_\pi$ bond. The above expression can be further simplified to $p_\sigma$-$p_\pi$, $p_\sigma$-$p_\sigma$ and $p_\pi$-$p_\pi$, leading to mediate, strong and weak interaction strengths, according to orbital wavefunction overlapping levels (Eq.7). Combining two super-exchange paths (Cr-Cl-Cr and Cr-O-Cr) with two bond angles (90º and 180º), there has four possible SEIs in CrOCl.

For Cr-Cl-Cr in 90º (Fig. 2a, path $P_1$ in Table 1), two $d^4$ states interact through two $p_\sigma$-$p_\pi$ bonds with mediate strength. Due to PEP, only in the condition that the spins of two cations are antiparallel, electron hoping between them is not forbidden, namely the interaction is AFM.

For Cr-O-Cr in 90º (Fig. 2b, path $P_2$), the interaction spreads through two $p_\sigma$-$p_\pi$ bonds while first $p_\sigma$-$p_\pi$ favors AFM according to PEP and the $p_\pi$-$p_\sigma$ bond favors FM due to HR that requires the maximum total magnetic moments. Since $d^3$ state has unfilled $e'_g$ orbital, HR takes advantage. In conclusion, this case is FM with extremely weak strength, owning to the competition between two $p_\sigma$-$p_\pi$ bonds.

For Cr-Cl-Cr in 180º (Fig. 2c, path $P_3$), the interaction conducted through one $p_\sigma$-

$p_\sigma$ bond and one $p_\pi$-$p_\pi$ bond. Both favor AFM (PEP). So, this case shall be AFM and extremely strong due to the strong character of $p_\sigma$-$p_\sigma$ bond.

For Cr-O-Cr in 180º angle (Fig. 2d, path $P_4$), the interaction happened through one $p_\sigma$-$p_\sigma$ bond that favors FM according to HR and one $p_\pi$-$p_\pi$ bond that is AFM thanks to PEP. As the interaction through $p_\pi$-$p_\pi$ is weak, this case $p_\sigma$-$p_\sigma$ bond would dominate which results in strong FM.

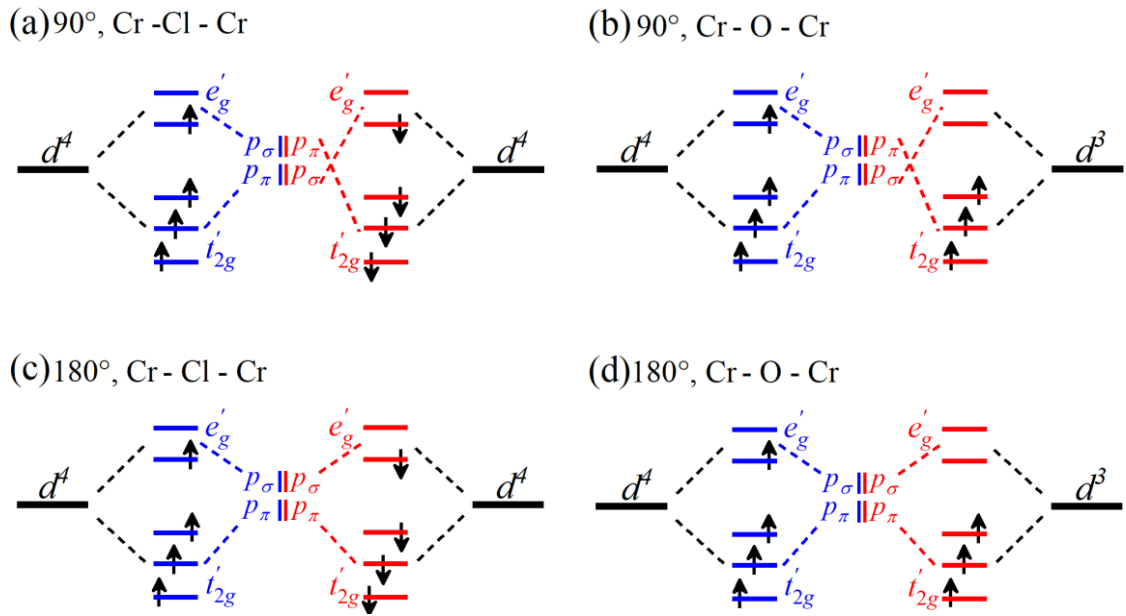

Fig. 2. Four Possible super-exchange paths in CrOCl. **(a, c)** Cr-Cl-Cr, 90º and 180º paths. **(b, d)** Cr-O-Cr, 90º and 180º paths. Electrons can only hop between orbitals that connected by dash blue and red lines.

Table 1. Predicted results of different paths in Fig. 2.

| Path | Interaction Path | State of Cation | Angle | Magnetic Ordering | Strength |
| --- | --- | --- | --- | --- | --- |
| $P_1$ | Cr-Cl-Cr | $\frac{1}{\sqrt{2}}(|d^4\rangle + |d^4\rangle)$ | 90º | AFM | Mediate |
| $P_2$ | Cr-O-Cr | $\frac{1}{\sqrt{2}}(|d^4\rangle + |d^3\rangle)$ | 90º | FM | Weak |
| $P_3$ | Cr-Cl-Cr | $\frac{1}{\sqrt{2}}(|d^4\rangle + |d^4\rangle)$ | 180º | AFM | Strong |
| $P_4$ | Cr-O-Cr | $\frac{1}{\sqrt{2}}(|d^4\rangle + |d^3\rangle)$ | 180º | FM | Strong |

The above predictions are automatically suitable for monolayer whose magnetic interactions are only through in-layer magnetic moments. However, for bulk CrOCl, from those evidence that (a) the exfoliation energy (Fig. S2) of monolayer is smaller than MnPSe$_3$[39] which possesses the lowest exfoliation energy as we know; (b) bulk is AA stacking; (c) experiment confirmed the magnetic moment orientation of bulk is in out-plane direction[22] (along z-axis), we can ignore the magnetic interaction between layers. So, predictions are suitable not only to monolayer, but also to bulk.

So far, ESET has fully predicted the magnetic orderings and strengths in different super-exchange paths. In real materials, magnetic exchange constants are directly related with those paths, which gives a effective way to verify our theory.

## 4 ESET Prediction Results with First-Principles Calculation

**4.1 Magnetic Exchange Constants**

In monolayer and bulk CrOCl, the SEI of cations has three different kinds that corresponding to first, second and third nearest neighbor magnetic exchange constants $J_1$, $J_2$ and $J_3$, as shown in Fig. 3a. To clearly see their super-exchange paths, we extracted them from the CrOCl material (Fig. 4). For exchange constants $J_1$, two Cr atoms interacted through two $P_2$ paths. Since $P_2$ are weak FM based on above analysis (as summarized in Table 1), $J_1$ shall also be weak FM. $J_2$ is through $P_1$ and $P_2$ paths. As $P_2$ AFM character is stronger than $P_1$ FM, $J_2$ is mediate AFM. For $J_3$, that one path $P_4$ makes it FM with strong strength. The results based on ESET are summarized in Table 2.

In order to get magnetic constants through first-principles calculation, we express the total energy of one magnetic atom for four configurations (Fig. 3) by Heisenberg Spin Hamiltonian[21]

$$E_{FM/AFM-Néel} = E_0 - (\pm 4J_1 + 2J_2 + 2J_3)|\vec{S}|^2$$

$$E_{AFM-stripy-x/y} = E_0 - (\mp 2J_2 \pm 2J_3)|\vec{S}|^2 \qquad (8)$$

where $E_0$ is the nonmagnetic total energy, $J_1$, $J_2$ and $J_3$ are magnetic exchange constants for first, second and third nearest neighbor and $|\vec{S}|$ is magnetic moment that equals to $3\mu_0$.

On the one hand, that both $J_1$ and $J_3$ are positive indicates their FM interaction while $J_2$ is AFM owing to its negative sign. On the other, the order of interaction strength is $|J_3| > |J_2| > |J_1|$. No matter in sign or magnitude, the calculation results are highly accordance with our ESET prediction results, which proves our theory.

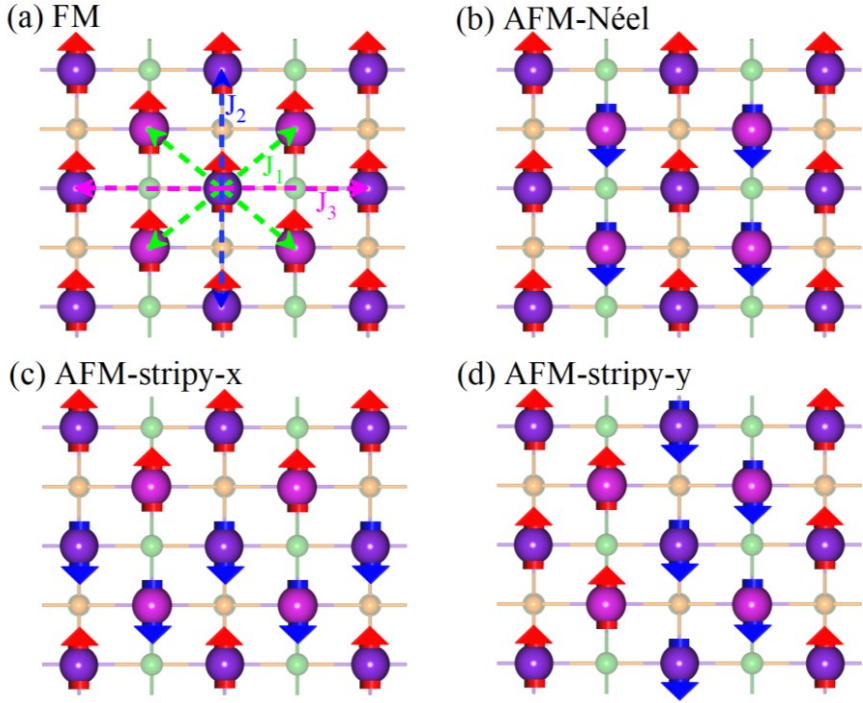

**Fig. 3. Four possible magnetic configurations of monolayer CrOCl and magnetic constants. (a) FM; (b) AFM-Néel; (c) AFM-Stripy-x; (d) AFM-Stripy-y.** The magnetic constants of first, second, third nearest neighbor are denoted as $J_1$, $J_2$ and $J_3$ in (a). The upper Cr atoms are shown in violet and lower Cr atoms are in pink.

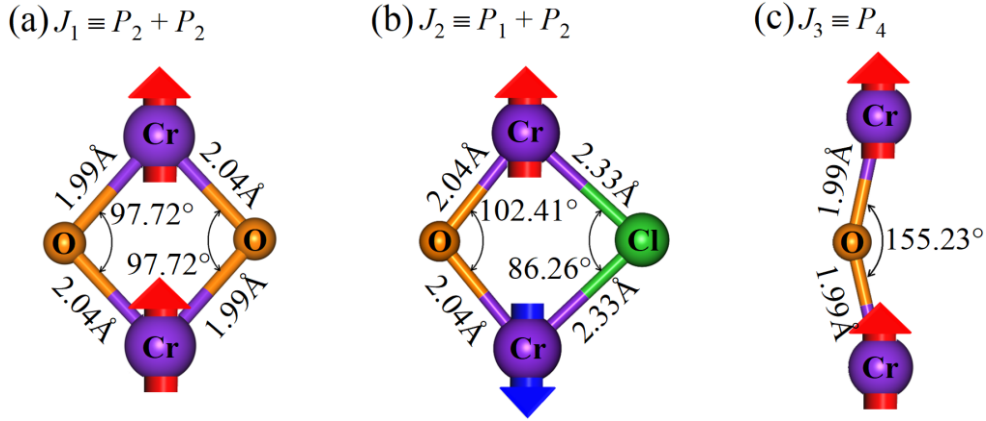

**Fig. 4. Super-exchange interaction paths (refer to Table 1) for magnetic exchange constants.** (a) $J_1$, including two $P_2$ paths; (b) $J_2$, including one $P_1$ path and one $P_2$ path; (c) $J_3$, including one $P_4$ path.

**Table 2. Predicted results from ESET and first-principles calculation of exchange constants in meV.** That prediction from our extended super-exchange theory and first-principles calculation are in good accordance verifies our theory.

|  | Interaction Path | Magnetic Ordering | Interaction Strength | Exchange Constants Monolayer | Bulk |
|---|---|---|---|---|---|
| $J_1$ | $P_2 + P_2$ | FM | Weak | 0.223 | 0.193 |
| $J_2$ | $P_1 + P_2$ | AFM | Mediate | - 0.391 | -0.363 |
| $J_3$ | $P_4$ | FM | Strong | 1.780 | 1.292 |

### 4.2 The magnetic ground state of CrOCl

Since the validity of ESET has been clarified by first-principles calculation, we now use it to predict the magnetic ground state of CrOCl based on Eq. (8).

The energy difference between $E_{FM}$ and $E_{AFM-stripy-x}$ ($E_{AFM-stripy-x} - E_{FM}$) equals to $4|S|^2(J_1 + J_2)$. Since $J_1$ is weak FM while $J_2$ is mediate AFM, it is obvious that $J_1 + J_2 < 0$. The energy of $E_{AFM-stripy-x}$ is smaller than $E_{FM}$, that is, the ground state of CrOCl should be AFM but energy difference is relatively small as dominated interaction $J_2$ is not a strong type. The first-principles calculation results give that $E_{AFM-stripy-y} - E_{FM}$ equals to -2.70 meV/Cr in monolayer and -3.06 meV/Cr in bulk. Besides, the experiment of AA stacking bulk CrOCl shown it is AFM at 4K and the

orientation of in-plane magnetic moments is corresponding to our AFM-stripy-x configuration [22]. Those evidences, again, verify the rationality of our theory.

By the way, we have noticed that conclusions from a recent paper[18] (monolayer CrOCl is FM ordering) are controversial to ours. Since their magnetic configurations are repeated, their calculation results need further verification.

## 5 Conclusion Remarks

In summary, we have come up with an extended super-exchange theory to predict the magnetic properties of materials that have polyvalent anions as the original super-exchange theory does not work anymore. The electronic states of cations in all super-exchange paths, in the framework of original super-exchange theory, are always the same. However, for polyvalent anion materials, due to the influence of different valence states anions, the electronic states of cation in different super-exchange paths must be taken good care and not identical. Based on CrOCl material, we elaborate the analytic procedures of extend super-exchange theory. We suggest, the electronic states of cations are superposed and depend on specific super-exchange path. Our theory can predict (a) the sign and relative magnitude of magnetic exchange constants for different cation-anion-cation paths; (b) overall magnetic ground state. Although we only demonstrated that ESET deals with *3d* orbital cations, expanding it to 4*d*, 5*d* and 4*f* orbitals are straightforward, which only requires modification of orbital symmetry relations. In addition, our theory reveals that SEI can lead to strong ferromagnetic ordering, while original theory denies its existence. From this point, high Curie temperature two-dimensional semiconducting ferromagnets are very likely to be discovered in polyvalent anion materials. Thus, our proposed ESET will have significant impact as more polyvalent anion materials are being studied.

## Associated Contents

**Supporting Information**

Calculation details and procedures for structure optimization, total energy, exfoliation energy and phonon-related properties; Magnetic properties of layered FeOCl analyzed by ESET.

Additional figures for PDOS picture of monolayer CrOCl with *AFM-stripy-x* magnetic configuration, exfoliation energy of monolayer CrOCl and other common 2D materials, phonon pictures; the scheme of four possible super-exchange paths in FeOCl.

Additional tables for crystal constants and total energies of layered CrOCl; ESET prediction and first-principles calculation results of magnetic exchange constants for layered FeOCl.

## Acknowledgments

This work in University of Macau is supported by the Start-up Research Grant (SRG2016-00002-FST), the Science and Technology Development Fund (FDCT 063/2016/A2) and the Research & Development Grant for Chair Professor (CPG2016-00026-FST). In Southern University of Science and Technology is supported by the Shenzhen Fundamental Research Foundation (Grant No. JCYJ20170817105007999), Natural Science Foundation of Guangdong Province of China (Grant No. 2017A030310661). In Zhengzhou University is supported by the National Natural Science Foundation Youth Fund (Grant No. 11704342).

**TOC**

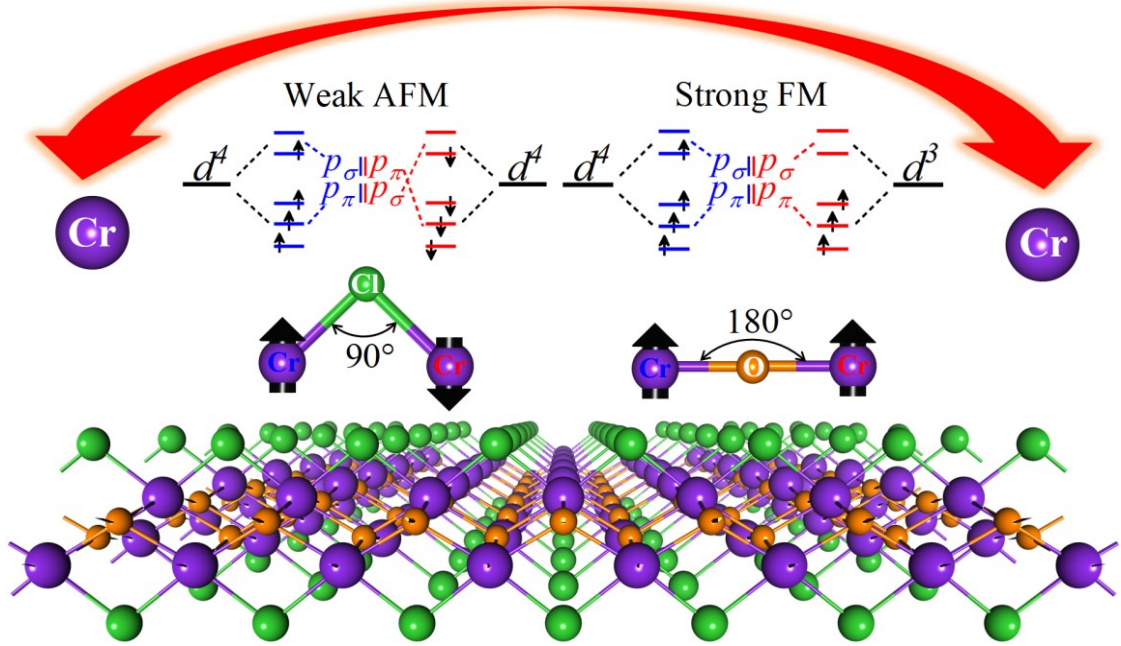